\documentclass[%
twocolumn%
,secnumarabic%
,amsfonts,amssymb,aps,prapplied,longbibliography,nobibnotes,superscriptaddress]{revtex4-1}
\usepackage{amsmath}
\usepackage{epsf,epstopdf,epsfig}
\usepackage{graphicx}
\usepackage{fancyhdr}
\usepackage{color}
\usepackage{xcolor}
\expandafter\ifx\csname package@font\endcsname\relax\else
 \expandafter\expandafter
 \expandafter\usepackage
 \expandafter\expandafter
 \expandafter{\csname package@font\endcsname}%
\fi
\DeclareRobustCommand\substyle{\name@idx{document substyle}}%
\DeclareRobustCommand\classoption{\name@idx{document class option}}%
\DeclareRobustCommand\classname{\name@idx{document class}}%
\def\name@idx#1#2{%
 {\ttfamily#2}%
 \index{#2\space#1=\string\ttt{#2}\space#1}\index{#1>#2=\string\ttt{#2}}%
}%

\AtBeginDocument{%
	\newwrite\bibnotes
	\def\bibnotesext{Notes.bib}
	\immediate\openout\bibnotes=\jobname\bibnotesext
	\immediate\write\bibnotes{@CONTROL{REVTEX41Control}}
	\immediate\write\bibnotes{@CONTROL{%
			apsrev41Control,author="08",editor="1",pages="1",title="0",year="1"}}
	\if@filesw
	\immediate\write\@auxout{\string\citation{apsrev41Control}}%
	\fi
}%

\begin{document}
\title{Rectification in Spin-Orbit Materials Using Low Energy Barrier Magnets}%

\author{Shehrin Sayed}\email{ssayed@berkeley.edu}
\affiliation{Electrical and Computer Engineering, Purdue
	University, West Lafayette, IN 47907, USA}
\affiliation{Electrical Engineering and Computer Science, University of California, Berkeley, CA 94720, USA}
\author{Kerem Y. Camsari}
\affiliation{Electrical and Computer Engineering, Purdue
	University, West Lafayette, IN 47907, USA}
\author{Rafatul Faria}
\affiliation{Electrical and Computer Engineering, Purdue
	University, West Lafayette, IN 47907, USA}
\author{Supriyo Datta}\email{datta@purdue.edu}
\affiliation{Electrical and Computer Engineering, Purdue
	University, West Lafayette, IN 47907, USA}


\begin{abstract}
The coupling of spin-orbit materials to high energy barrier ($\sim$40-60 $k_BT$) nano-magnets has attracted growing interest for exciting new physics and various spintronic applications. We predict that a coupling between the spin-momentum locking (SML) observed in spin-orbit materials and low-energy barrier magnets (LBM) should exhibit a unique multi-terminal rectification for arbitrarily small amplitude channel currents. The basic idea is to measure the charge current induced spin accumulation in the SML channel in the form of a magnetization dependent voltage using an LBM, either with an in-plane or perpendicular anisotropy (IMA or PMA). The LBM feels an instantaneous spin-orbit torque due to the accumulated spins in the channel which causes the average magnetization to follow the current, leading to the non-linear rectification. We discuss the frequency band of this multi-terminal rectification which can be understood in terms of the angular momentum conservation in the LBM. For a fixed spin-current from the SML channel, the frequency band is same for LBMs with IMA and PMA, as long as they have the same total magnetic moment in a given volume. The proposed all-metallic structure could find application as highly sensitive passive rf detectors and as energy harvesters from weak ambient sources where standard technologies may not operate.
\end{abstract}
\maketitle

\section{Introduction}

The interplay between spin-orbit materials and nano-magnetism has attracted much attention for interesting phenomena e.g. spin-orbit torque switching \cite{Ralph_Science_2012, Ralph_PRL_2012}, probing the spin-momentum locking \cite{JonkerNatNano2014, ChenPRL2017, SamarthPRB2015,Dash_NanoLett_2015, Pham_NanoLett_2016, Koo_New}, spin amplification \cite{AGhosh_PRL_2015}, spin battery \cite{ChenScience2017}, skyrmion dynamics \cite{Woo2017, Wang_NanoLett2016}, among other examples. In this paper, we predict that the spin-momentum locking (SML) observed in spin-orbit materials when coupled to a nano-magnet with low-energy barrier, will rectify the channel current in the form of a voltage in a multi-terminal structure. We start our arguments with the spin-potentiometric measurements well-established in diverse classes of spin-orbit materials (see, for example, \cite{JonkerNatNano2014, ChenPRL2017, SamarthPRB2015, Dash_NanoLett_2015, Pham_NanoLett_2016, Koo_New}) where a high-energy-barrier stable ferromagnet (FM) is used to measure the charge current induced spin potential in the SML channel. We show that such spin-potential measurement on a metallic SML channel using a low-energy barrier magnet (LBM) will result in a rectified voltage, even for arbitrarily small channel current. 

The discussions on the multi-terminal rectification is limited in the linear response regime of transport in the SML channel and the non-linearity occurs due to the spin-orbit torque (SOT) driven magnetization dynamics of the LBM. We further show that the multi-terminal rectification is limited by a characteristic frequency of the LBM that can be understood in terms of angular momentum conservation between the spins injected from spin-orbit materials and the spins absorbed by the LBM. We argue that, for a fixed spin-current from the SML channel, the characteristic frequency is the same for LBMs with in-plane and perpendicular magnetic anisotropies (IMA and PMA), as long as they have the same total magnetic moment in a given volume.

We analyze the rectification in the proposed all-metallic structure (see Fig. \ref{1}(a)) considering both IMA and PMA LBMs and provide simple models to understand the underlying mechanisms of (i) the spin-orbit torque (SOT) induced magnetization pinning and (ii) the frequency band of the rectification. We compare the simple models with detailed numerical simulations using an experimentally benchmarked multi-physics framework \cite{Camsari_SciRep_2015}. The simulations are carried out using a transmission line model for the SML channel \cite{Sayed_arXiv_2017} and a stochastic Landau-Lifshitz-Gilbert (s-LLG) model for LBM \cite{PhysRevX.7.031014}, considering thermal noise within the magnet. We consider the spin-polarization axis to be in-plane of the SML channel and orthogonal to the current flow direction. Hence, in the present discussion, pinning for a IMA or a PMA magnet occurs along the easy-axis or hard-axis, respectively.

We argue that such wideband rectification in an all-metallic structure (Fig. \ref{1}(a)) could be used for `passive' radio frequency (rf) detection. Recently, Magnetic Tunnel Junction (MTJ) diodes with stable magnet as free layer and under an external dc current bias have demonstrated orders of magnitude higher sensitivity compared to the state-of-the-art Schottky diodes \cite{Miwa2013, Fang2016, doi:10.1063/1.5047547}. However, the reported no-bias sensitivity is lower or comparable to that of semiconductor diodes. The low-barrier nature of the magnet in the proposed structure should exhibit no-bias sensitivity as high as those observed using state-of-the-art technologies under external biases \cite{Miwa2013, Fang2016, doi:10.1063/1.5047547}. Furthermore, we discuss the possibility to harvest energy from weak ambient sources where standard technologies may not operate.

The paper is organized as follows. In Section \ref{sec2}, we establish the concept of the multi-terminal rectification in the SML channel using LBM, starting from the well-established spin-potentiometric measurements typically done with high-energy-barrier stable magnets. In Section  \ref{sec3}, we discuss the frequency bandwidth of the rectification and provide a simple model that applies to LBMs with both IMA and PMA. We argue using detailed simulaiton results that such bandwidth arises due to the principles of angular momentum conservation between the spins injected from the SML channel and the spins absorbed by the LBM. In Section IV, we discuss possible applications of the proposed all-metallic structure in `passive' rf detection and energy harvesting. We argue that the no-bias sensitivity of the proposed rectifier can be as high as those observed in state-of-the-art technologies under external bias. Finally, in Section \ref{sec5}, we end with a brief conclusion.
 
\section{Multi-Terminal Rectification}
\label{sec2}
\begin{figure}
	\includegraphics[width=0.48 \textwidth]{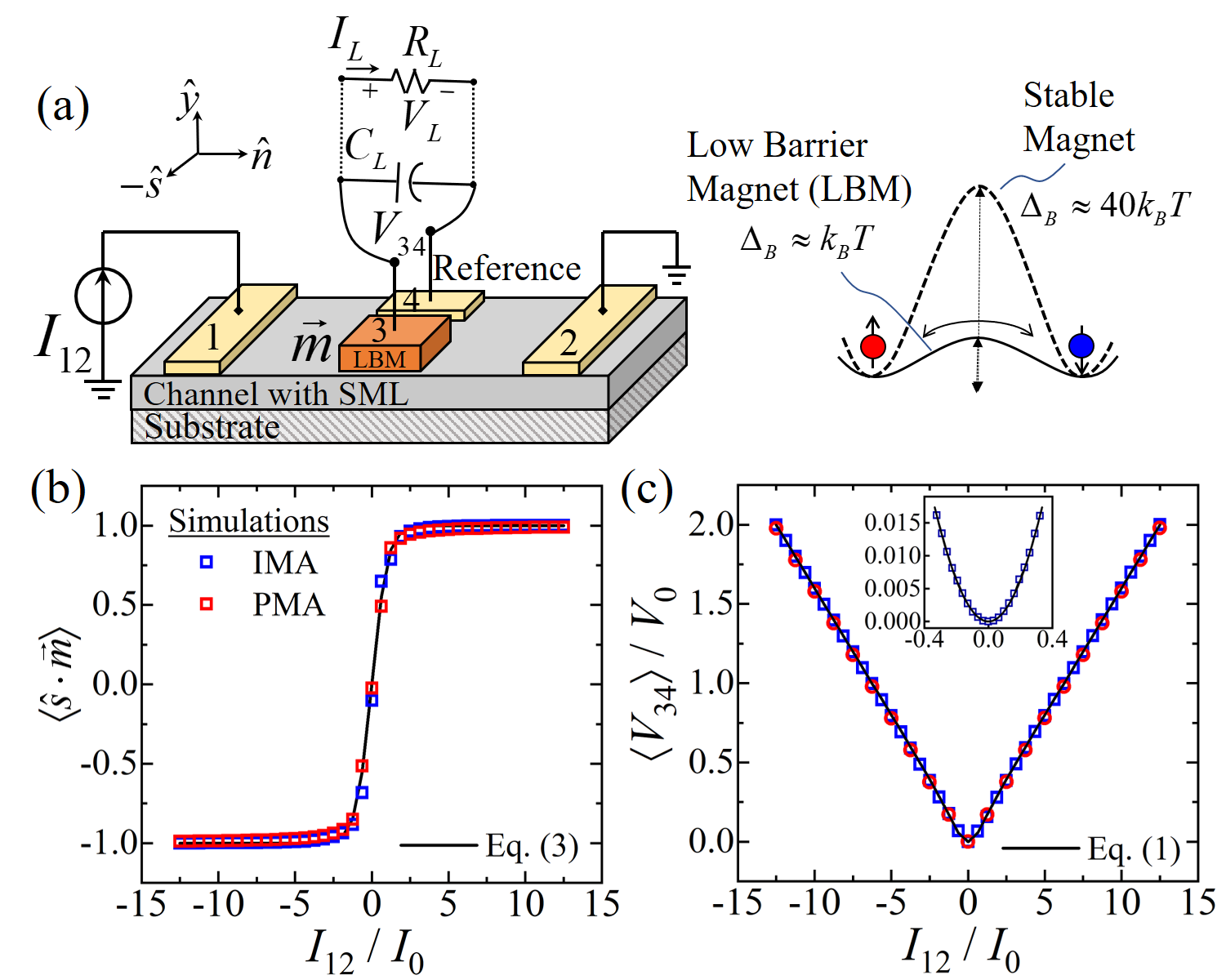}
	\caption{(a) Multi-terminal structure with a low barrier magnet (LBM) on top of a channel with spin-momentum locking (SML). LBM can be of in-plane and perpendicular anisotropies (IMA or PMA). (b) Average magnetization $\langle \hat{s}\cdot\vec{m} \rangle$ of the LBM. (c) Average voltage $\langle V_{34} \rangle$ between the LBM and a reference contact as a function of the input current $I_{12}$. Inset shows zoomed version of $\langle V_{34} \rangle$ for very small input current, which exhibits a parabolic nature. Simulations are compared with Eqs. \eqref{spin_vol} and \eqref{avg_Vfm}. $I_0$ for IMA and PMA are 80 $\mu$A and 1.6 $\mu$A respectively. Here, $V_0=I_0 R_B$.}\label{1}
\end{figure}

We start our arguments with the well-established spin-potentiometric measurements \cite{JonkerNatNano2014, ChenPRL2017, SamarthPRB2015, Dash_NanoLett_2015, Pham_NanoLett_2016, Koo_New} where the charge current induced spin potential in the SML channel is measured in the form of a magnetization dependent voltage using a stable ferromagnet (FM). The voltage at the FM with respect to a reference normal metal (NM) contact, placed at the same position along the current path as the FM (see Fig. \ref{1}(a)), is given by \cite{Hong_PRB_2012, SayedEDL2017}
\begin{equation}
\label{spin_vol}
{V_{34}\left(\vec m\right)} = {\left(\hat{s}\cdot \vec{m} \right)}\,\dfrac{{\alpha \xi {p_0}{p_f}R_B}}{{2}}{I_{12}},
\end{equation}
which shows opposite signs for the two magnetic states of the FM under a fixed channel current $I_{12}$ flowing along $\hat{n}$-direction (see Fig. \ref{1}(a)). Here $\hat{s}$ is the spin polarization axis in the SML channel defined by $\hat{y}\times \hat{n}$ with $\hat{y}$ being the out-of-plane direction \cite{Sayed_arXiv_2017}, $\vec{m}$ is the magnetization vector, $p_f$ is the FM polarization, $0\leq\xi\leq1$ is the current shunting factor \cite{Sayed_SciRep_2016} of the contact with 0 and 1 indicating very high and very low shunting respectively, $p_0$ is the degree of SML in the channel \cite{Sayed_arXiv_2017}, $\alpha\approx 2/\pi$ is an angular averaging factor \cite{Sayed_arXiv_2017}, and $R_B=(h/q^2)(1/M_t)$ is the ballistic resistance of the channel with total number of modes $M_t$ ($q:$ electron charge, $h:$ Planck's constant).

Note that Eq. \eqref{spin_vol} is valid all the way from ballistic to diffusive regime of operation \cite{Hong_PRB_2012, Sayed_arXiv_2017}. We restrict our discussion to linear response where ${V_{34}\left(\vec m\right)}$ in Eq. \eqref{spin_vol} scales linearly with $I_{12}$ and satisfies the Onsager reciprocity relation \cite{Buttiker_PRB_2012, Sayed_SciRep_2016} 
\begin{equation*}
R_{ij,kl}\left(\vec m\right) = R_{kl,ij}\left(-\vec m\right),
\end{equation*}
with $R_{ij,kl}=V_{kl}/I_{ij}$. The Onsager reciprocity does not require any specific relation between $R_{ij,kl}\left(\vec m\right)$ and $R_{ij,kl}\left(-\vec m\right)$ in linear response and the phenomenon described by Eq. \eqref{spin_vol} has been observed on diverse spin-orbit materials e.g. topological insulator (TI) \cite{JonkerNatNano2014, ChenPRL2017, SamarthPRB2015, Dash_NanoLett_2015}, Kondo insulators \cite{Hong_KIST_arXiv_2018}, transition metals \cite{Pham_NanoLett_2016}, semimetals \cite{Li_NatComm_2018}, and semiconductors \cite{Koo_New}.

To measure Eq. \eqref{spin_vol} from a highly resistive SML channel (e.g. TI  \cite{JonkerNatNano2014, ChenPRL2017, SamarthPRB2015,Dash_NanoLett_2015}, semiconductor \cite{Koo_New}, etc.)  using a metallic FM, usually a thin tunnel barrier is inserted at the interface. This tunnel barrier effectively enhances $V_{34}$ by improving $\xi$ \cite{Sayed_SciRep_2016}, however, degrades the spin injection into the FM from the SML channel. It has been recently demonstrated \cite{Pham_NanoLett_2016} that $V_{34}$ can be measured with metallic FM in direct contact with metallic SML channels (e.g. Pt, Ta, W, etc.), which indicates the possibility of spin-voltage reading (e.g. \cite{Pham_NanoLett_2016}) and spin-orbit torque (SOT) writing (e.g. \cite{Ralph_Science_2012, Ralph_PRL_2012}) of the nano-magnet within same setup with different current magnitudes \cite{SayedEDL2017}.

The energy barrier of a mono-domain magnet is given by $\Delta_B=\frac{1}{2}H_k M_s \Omega$ \cite{SunPRB2000} where $H_k$ is the anisotropy field, $M_s$ is the saturation magnetization, and $\Omega$ is the FM volume. For a stable FM, $\Delta_B \approx 40\sim60\;k_BT$ and exhibit very long retention time $\tau \propto \exp\left(\Delta_B/k_BT\right)$ of the magnetization state ($k_B:$ Boltzmann constant, $T:$ temperature). LBMs have very small $\tau$ and the $\hat{s}\cdot\vec{m}$ component becomes random within the range $\left\{+1,-1\right\}$ driven by the thermal noise. Experimentally, LBMs have been achieved by lowering the total moment ($M_s\Omega$) \cite{PhysRevApplied.8.054045} or by lowering the anisotropy field ($H_k$) either by increasing the thickness of a PMA \cite{8421599}, or by making a circular IMA with no shape anisotropy \cite{7838539}.

At equilibrium ($I_{12}=0$), the time-averaged $\langle \hat{s}\cdot\vec{m} \rangle = 0$ for an LBM. For $I_{12}\neq 0$, induced non-equilibrium spins in the channel apply SOT on the LBM and  $\langle \hat{s}\cdot\vec{m} \rangle$ follows the accumulated spins, which can be calculated using
\begin{equation}
\label{av_fpe1}
\langle \hat{s}\cdot\vec{m} \rangle = \dfrac{\displaystyle
	\int_{\phi=-\pi}^{\phi=\pi}\displaystyle
	\int_{\theta=0}^{\theta=\pi} \left(\hat{s}\cdot\vec{m}\right) \;\rho\;\sin\theta\;d\theta d\phi}{\displaystyle
	\int_{\phi=-\pi}^{\phi=\pi}\displaystyle
	\int_{\theta=0}^{\theta=\pi}\rho\;\sin\theta\;d\theta d\phi}.
\end{equation}
where $\rho$ is the probability distribution function of the magnetization of the LBM under a particular $I_{12}$, which can be obtained from the Fokker-Planck equation \cite{PhysRev.130.1677, ButlerFPE2012}. The dependence of $\langle \hat{s}\cdot\vec{m} \rangle$ on $I_{12}$ deduced from Eq. \eqref{av_fpe1} for a particular LBM, can in-principle be any saturating odd-functions e.g. Langevin function  for low-barrier PMA (see Appendix \ref{App_I_0}).

We approximate $\langle \hat{s}\cdot\vec{m} \rangle$ in Eq. \eqref{av_fpe1} with a $\tanh$ functional dependence on $I_{12}$, given by
\begin{equation}
\label{avg_Vfm}
\langle \hat{s}\cdot\vec{m} \rangle\approx \tanh\left(\dfrac{I_{12}}{I_0}\right),
\end{equation}
which is in  good agreement with the detailed numerical simulations for both IMA and PMA, as shown in Fig. \ref{1}(b). The simulations are carried out within a multi-physics framework \cite{Camsari_SciRep_2015} using our experimentally benchmarked transmission line model for SML \cite{Sayed_arXiv_2017} and stochastic Landau-Lifshitz-Gilbert (s-LLG) model for LBM \cite{PhysRevX.7.031014} which considers thermal noise. The details of the simulation setup is discussed in Appendix \ref{App_Sim}.

\begin{figure*}
	\includegraphics[width=1 \textwidth]{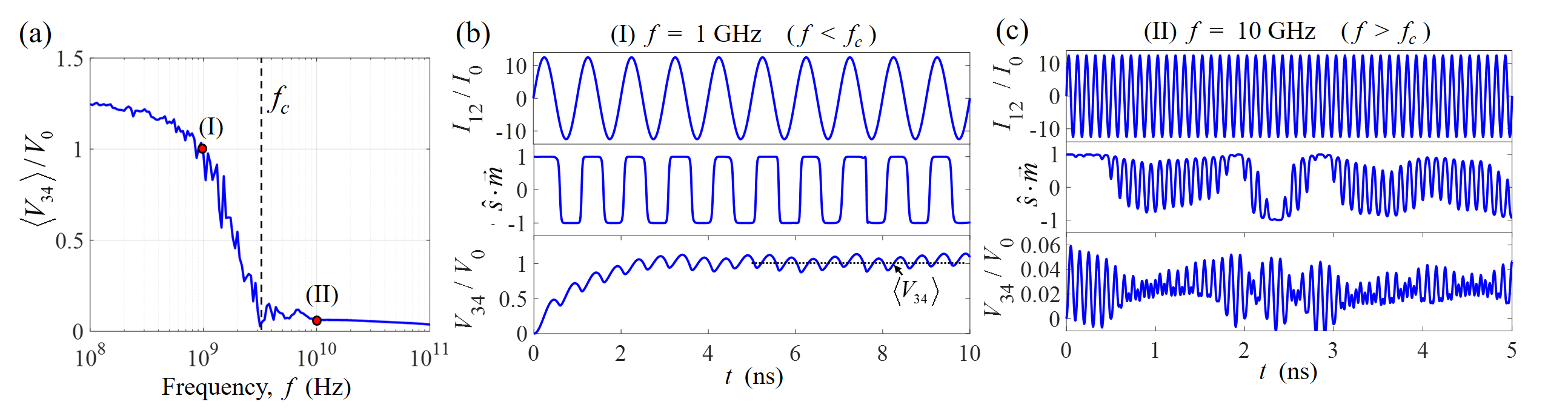}
	\caption{(a) Rectified voltage $\langle V_{34} \rangle$ as a function of the input ac frequency $f$, showing the frequency bandwidth $f_c$. We observe the time dynamics of the LBM under input current $I_{12}$ for (b) $f<f_c$, where $\hat s \cdot \vec m$ on average follows $I_{12}$ leading to rectification and (c) $f>f_c$, where $\hat s \cdot \vec m$ struggles to follow $I_{12}$ and yields no net rectification. Results apply to both IMA and PMA. Here, we consider 49 nm $\times$ 61 nm $\times$ 5 nm LBM with $M_s=900$ emu/cc. $i_{s0}=\beta i_{c0} = 2$ mA.
	}\label{sup2}
\end{figure*}

Here, $I_0$ is a parameter that determines the SOT induced magnetization pinning of the LBM. $I_0$ depends on the temperature, geometry, and material parameters and much larger for an IMA as compared to a PMA due to the demagnetization field. $I_{12}$ along $\mp\hat{n}$-direction causes the magnetization pinning along $\pm\hat{s}$-direction. In the present discussion, easy axis for PMA is along $\hat y$-direction, hence, the pinning occurs along the hard axis. The easy axis of IMA, in principle, can be in any direction on the plane spanned by $\hat{n}$ and $\hat{s}$ and the magnetization pinning along $\hat{s}$-direction should be described by Eq. \eqref{avg_Vfm} with a modified $I_0$. However, we set the easy axis along $\pm\hat{s}$-direction in our IMA simulations for simplicity.

For a given structure, $I_0$ can be determined directly from experiments using a characteristic curve similar to that in Figs. \ref{1}(b) or (c). We provide a simple expression using Eq. \eqref{av_fpe1} and considering easy-axis pinning of a PMA magnet (see Appendix \ref{App_I_0} for the derivation), as given by
\begin{equation}
\label{I0_LBM}
{I_0} \approx \frac{{6q}}{\hbar }\frac{{{k_B}T{\alpha _g}}}{\beta},
\end{equation}
where $\alpha_g$ is the Gilbert damping and $\beta$ is the charge to spin current conversion ratio. Eq. \eqref{I0_LBM} is reasonably valid up to $\Delta_B\approx k_BT$ and provides the correct order of magnitude up to several $k_BT$ (see Appendix \ref{App_I_0} for details). In this discussion, we consider very low energy barrier ($\leq 1k_BT$) nano-magnets that do not have bistable states. A higher barrier magnet that exhibits bistable states, in principle could exhibit effects like stochastic resonance \cite{PhysRevLett.105.047202}, which is not the subject of the present discussion.

For $|I_{12}|\gg I_0$ in Eq. \eqref{avg_Vfm}, we have $\tanh\left(I_{12}/I_0\right)\approx$ $+1$ or $-1$ when $I_{12}>0$ or $I_{12}<0$ respectively. Hence, $I_{12}\times\tanh\left(I_{12}/I_0\right)\approx |I_{12}|$. On the other hand, for $|I_{12}|\ll I_0$ in Eq. \eqref{avg_Vfm}, we have $I_{12}\times\tanh\left(I_{12}/I_0\right)\approx I_{12}^2/I_0$. Thus, the time-average of the voltage in Eq. \eqref{spin_vol} is given by
\begin{equation}
\label{avg_spin_vol}
\langle V_{34} \rangle=
\begin{cases}
\left(\dfrac{{\alpha \xi {p_0}{p_f}R_B}}{{2I_0}}\right)\,{I_{12}^2}, & \text{for $|I_{12}| \ll I_0$} \\
\left(\dfrac{{\alpha \xi {p_0}{p_f}R_B}}{{2}}\right)\,{|I_{12}|}. & \text{for $|I_{12}| \gg I_0$}
\end{cases}
\end{equation}
Note that $\langle V_{34} \rangle$ represents the steady-state voltage of the capacitor $C_L$ placed between contacts 3 and 4. The relative position between contacts 3 and 4 along $\hat{s}$-direction do not affect Eq. \eqref{spin_vol}, however, a shift in the $\hat{n}$-direction creates an offset due to Ohmic drop \cite{SayedEDL2017}, that should cancel out over averaging in Eq. \eqref{avg_spin_vol} when an ac $I_{12}$ is applied. For arbitrary $I_{12}$, $\langle V_{34} \rangle$ is always of the same sign leading to a multi-terminal rectification. This observation agrees well with simulation results for IMA and PMA, as shown in Fig. \ref{1}(c). For $|I_{12}|\ll I_0$, $\langle V_{34} \rangle$ exhibits a parabolic nature (see zoomed inset of Fig. \ref{1}(c)), as suggested by Eq. \eqref{avg_spin_vol}. All simulation results presented in this paper are normalized by $I_0$, $V_0=I_0R_B$, and $f_0=I_0/q$ for current, voltage, and frequency, respectively. In all simulations, $M_t=100$ which yields $R_B=259\Omega$.

\section{Frequency Bandwidth}
\label{sec3}

The frequency bandwidth of the the multi-terminal rectification is limited by a characteristic frequency $f_c$ that is determined by the angular momentum conservation between the spins injected from the SML channel and the spins absorbed by the LBM. We plot the $\langle V_{34} \rangle$ as a function of the frequency $f$ of the ac $I_{12}=i_{c0}\,\sin\left(2\pi f t \right)$ (see Fig. \ref{sup2}(a)) while other parameters are kept constant in our simulations. Note that $\langle V_{34} \rangle$ is relatively constant in the low frequency region and degrades significantly for $f>f_c$.  We have defined $f_c$ as the frequency where $\langle V_{34} \rangle$ degrades by an order of magnitude compared to the region where $\langle V_{34} \rangle$ vs. $f$ is relatively flat.

We observe the time-dynamics of $\hat{s}\cdot \vec{m}$ and $V_{34}$ for the two cases indicated with red-dots in Fig. \ref{sup2}(a): (I) $f<f_c$ and (II) $f>f_c$. For the first case, $I_{12}$ is slow enough that the injected spins from SML channel into the LBM satisfies the angular momentum conservation and $\hat{s}\cdot \vec{m}$ follows the $I_{12}$ at the same frequency, as shown in Fig. \ref{sup2}(b). This leads to a rectified voltage $V_{34}$ that charges up the capacitor $C_L$ to the steady-state value $\langle V_{34} \rangle$. The ripples observed in $V_{34}$ is similar to those in conventional rectifiers and gets attenuated for increased $C_L$. For the latter case, $\hat{s}\cdot \vec{m}$ struggles to follow $I_{12}$ (see Fig. \ref{sup2}(c)) since the spins injected from the SML channel to the LBM is fast enough that they do not satisfy the angular momentum conservation. $\hat{s}\cdot \vec{m}$ has no correlation with $I_{12}$, as a result, there is no rectification that charges up $C_L$ to a steady dc voltage.

We obtain an empirical expression for $f_c$ from the detailed s-LLG simulations using a broad range of parameter values, given by
\begin{equation}
\label{cut_off}
2\pi {f_c} = \dfrac{i_{s0}}{{2q{N_s}}},
\end{equation}
where the injected spin current amplitude $i_{s0}=\beta i_{c0}$, $N_s=M_s \Omega/\mu_B$ is the total number of spins in LBM and $\mu_B$ is the Bohr magneton. The functional dependence of $f_c$ on $i_{s0}$ and $N_s$ is very similar to the switching delay for stable magnets \cite{Behtash2011} that also arises from the principles of angular momentum conservation. Note that Eq. \eqref{cut_off} is valid for both IMA and PMA.

\begin{figure}
	\includegraphics[width=0.42 \textwidth]{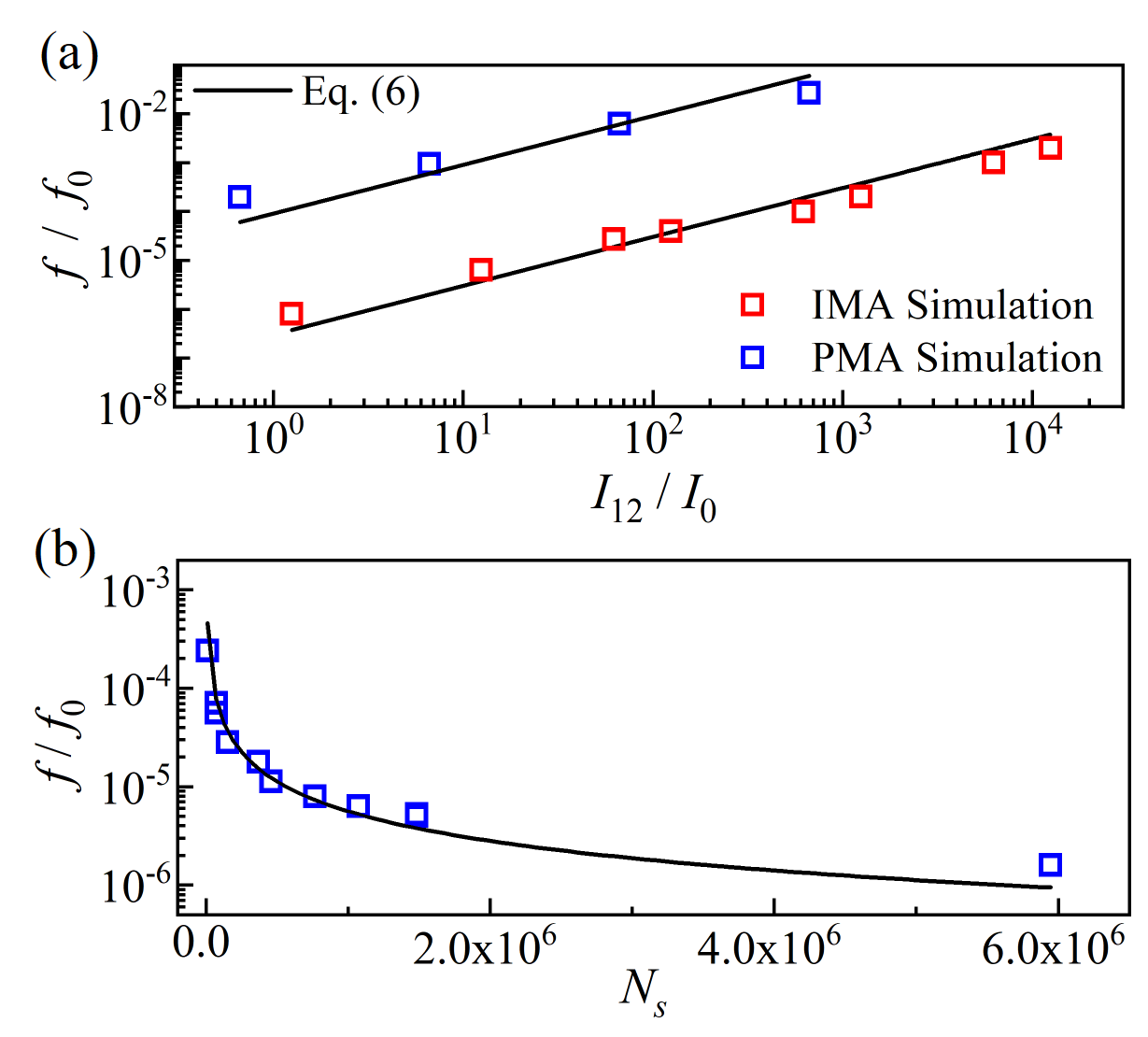}
	\caption{Characteristic frequency $f_c$ is (a) proportional to $i_{c0}$ ($N_s=$ $10^4$ and $10^6$ for PMA and IMA respectively) and (b) inversely proportional to $N_s$. Here, $f_0=I_0/q$.
	}\label{4}
\end{figure}

We show comparison between Eq. \eqref{cut_off} and simulation results in Fig. \ref{4}. The simulation data points shown on Fig. \ref{4} are extracted from a plot similar to Fig. \ref{sup2}(a). Eq. \eqref{cut_off} shows good agreement with the simulation for LBMs having IMA with easy-axis pinning and PMA with hard-axis pinning (see Figs. \ref{4}(a)-(b)). Fig. \ref{4}(a) shows that $f_c$ scales linearly with $i_{s0}$. A similar scenario has been reported \cite{PhysRevApplied.2.034009} for a stochastic MTJ oscillator made with relatively lower-barrier free magnetic layer. According to Eq. \eqref{cut_off} and detailed simulations, the conclusion that $f_c\propto i_{s0}$ seems valid even if $i_{s0}$ changes by orders of magnitude. Moreover, Fig. \ref{4}(b) shows that $f_c$ scales inversely proportional to the $N_s$ which depends only on $M_s\Omega$ of the magnet and independent of the magnetic anisotropy. Eq. \eqref{cut_off} could be useful for recent interest on LBM based applications e.g. stochastic oscillators \cite{PhysRevApplied.2.034009}, random number generators \cite{doi:10.1063/1.5006422,PhysRevApplied.8.054045}, probabilistic spin logic \cite{PhysRevX.7.031014, Mizrahi2018}, etc.

\section{Applications: RF Detection and Energy Harvesting}
\label{sec4}

The proposed all-metallic structure can find useful applications like rf detection and energy harvesting. In this section, we show that the low-barrier nature of the magnet can lead to very high rf detection sensitivity without any external bias, comparable to those observed in state-of-the-art technologies under an external bias. We provide a simple model for no-bias sensitivity which provides insight into the design of a high-sensitivity device. This could be of interest for rf detection from weak sources typically proposed to sense with quantum sensors (see, e.g., \cite{House2015,Gely1072}).We further argue that the proposed structure can extract useful energy from the ambient rf energy, especially from the weak sources where standard technologies may not operate.

It can be seen from Eq. \eqref{avg_spin_vol} that $\langle V_{34} \rangle$ scales  $\propto \sqrt{P_{in}}$ when $\max\left(I_{12}\right) \gg I_0$, where $P_{in} =\frac{1}{T}\int_{0}^{T}I_{12}^2\left(t\right) R_{12}\,dt$, $T=1/f$, and $R_{12}$ is the channel resistance. However, for $\max\left(I_{12}\right) \ll I_0$, $\langle V_{34} \rangle$ scales $\propto P_{in}$ with a constant slope given by
\begin{equation}
\label{sensitivity}
\dfrac{d\langle V_{34} \rangle}{dP_{in}}  = \dfrac{{\alpha \xi {p_0}{p_f}}}{{2}}\dfrac{{R_B}}{{{I_0}{R_{12}}}}.
\end{equation}
The derivation is given in Appendix \ref{App_ACtoDC}. The quantity in Eq. \eqref{sensitivity} is often considered as the sensitivity of rf detectors \cite{Miwa2013, Fang2016, doi:10.1063/1.5047547}. Recently, Magnetic Tunnel Junction (MTJ) diodes with stable magnet as free layer and under an external dc current bias have demonstrated orders of magnitude higher sensitivity compared to the state-of-the-art Schottky diodes \cite{Miwa2013, Fang2016, doi:10.1063/1.5047547}. However, the reported no-bias sensitivity is lower or comparable to that of semiconductor diodes. Eq. \eqref{sensitivity} indicates that the no-external-bias sensitivity can be very high within the all-metallic structure in Fig. \ref{1}(a) when designed to have very low $I_0$, enabling highly sensitive `passive' rf detection. With $\alpha_g=0.01$ and $T=300$ K we have $I_0 \approx 0.37\mu$A$/\beta$ from Eq. \eqref{I0_LBM}. For a Py LBM of dimension of 49 nm $\times$ 61 nm $\times$ 5 nm (see Ref. \cite{8421599}) and 2 nm thick Pt channel, $\beta$ can be $\sim 2$ as estimated from the charge to spin conversion ratio reported in Ref. \cite{Ralph_PRL_2012}, yielding $I_0\approx0.18$ $\mu$A. Note that $\beta$ can be much higher based on the geometry and the choice of the SML material.

For Bi$_2$Se$_3$ and Pt, we roughly estimate the sensitivity as 21,000 and 860 mV/mW respectively, assuming 2D SML channel of width $w=$ 210 nm and length $L=$ 500 nm. These estimations were done based on Eq. \eqref{sensitivity} using: (i) $R_B=$ 259 $\Omega$ (Bi$_2$Se$_3$) and 58 $\Omega$ (Pt), (ii) $R_{ch}\approx$ $6.5$ k$\Omega$ (Bi$_2$Se$_3$) and $\sim3$ k$\Omega$ (Pt), (iii) $p_0\approx$ 0.6 (Bi$_2$Se$_3$) and 0.05 (Pt) (see Ref. \cite{SayedEDL2017}), and (iv) $p_f\approx0.5$ \cite{Koo_New}. We have assumed $\xi\approx 1$ and the quoted estimations will be lower for higher shunting. $R_B$ has been estimated using $M_t=k_F w/\pi$, where $k_F=1.5$ nm$^{-1}$ (Bi$_2$Se$_3$) and $6.7$ nm$^{-1}$ (Pt) \cite{SayedEDL2017}. The channel resistance has been estimated using $R_{ch}=R_B(L+\lambda)/\lambda$ with mean free path $\lambda$ of 20 nm (Bi$_2$Se$_3$ \cite{Wang2016}) and 10 nm (Pt \cite{Johann_PRB_1980}), respectively. More detailed analysis and performance evaluation considering signal-to-noise ratio we leave for future work.

With proper materials and geometry, it may be possible to extract usable energy from such rectification of rf signals, especially from weak ambient sources. The dc power $P_L=V_LI_L$ extracted by an arbitrary load $R_L$ is limited by the equivalent resistance $R_{34}$ between contacts 3 and 4. The maximum efficiency of such rf to dc power conversion occurs for $\max\left(I_{12}\right) \gg I_0$, given by
\begin{equation}
\label{efficiency}
\eta_{\max}=\dfrac{P_{L,\max}}{P_{in}}=\left(\dfrac{{\alpha \xi {p_0}{p_f}}}{\sqrt{2}{\pi}}\right)^2  \dfrac{R_B}{{R_{12}}}\dfrac{R_B}{{R_{34}}}.
\end{equation}
The derivation is given in Appendix \ref{App_Efficiency}. Note that the maximum efficiency is independent of $I_0$. Assuming $R_{eq}=10R_B$ for enhanced $\xi$, we estimate the maximum efficiency to be 0.001\% for Bi$_2$Se$_3$ and 3$\times$10$^{-6}$\% for Pt even with $P_{in}$ in the $\sim$pW range given $I_0\leq0.18\mu$A. MTJ diodes recently demonstrated rf energy harvesting with similar efficiency \cite{WANG_arXiv_2018}, however, the input power was in the $\mu$W range. Such MTJs should achieve reasonable efficiency at lower input power if the stable free layer is replaced with an LBM.



\section{Conclusion}
\label{sec5}
In conclusion, we predict multi-terminal rectification in an all-metallic structure that comprises a spin-orbit material exhibiting spin-momentum locking (SML) and a low-energy barrier magnet (LBM) having either in-plane or perpendicular anisotropy (IMA or PMA). The discussion of such multi-terminal rectification was limited in the linear response regime of transport and the non-linearity occurs due to the spin-orbit torque driven magnetization dynamics of the LBM. We draw attention to a frequency band of the rectification which can be understood in terms of angular momentum conservation within the LBM. For a fixed spin-current from the SML channel, the frequency band is same for LBMs with IMA and PMA, as long as they have the same total magnetic moment for a given volume. We further discuss possible applications of the wideband rectification as highly sensitive passive rf detectors and as energy harvesters from ambient sources.

\begin{acknowledgments}
	This work was supported by ASCENT, one of six centers in JUMP, a SRC program sponsored by DARPA.
\end{acknowledgments}

\appendix
\section{Average Magnetization of Low-Barrier Magnets and Magnetization Pinning Current}
\label{App_I_0}
\noindent\textit{This section discusses the pinning current of a LBM and derives Eqs. \eqref{avg_Vfm}-\eqref{I0_LBM}, starting from the steady-state solution of the Fokker-Planck Equation.}\\

We start from the steady-state solution of probability distribution from Fokker-Planck Equation assuming a magnet with perpendicular magnetic anisotropy (PMA) (see Eq. (4.3) in Ref. \cite{ButlerFPE2012}), given by
\begin{equation}
\label{FPE}
\rho\left(m_z\right)=\dfrac{1}{{\rm Z}}\exp\left[ -\dfrac{\Delta_B}{k_B T}\left(1-m_z^2+2\left(\dfrac{H_{ext}}{H_k}+\dfrac{i_s}{I_{s0}}\right)m_z\right) \right],
\end{equation}
where $\rm Z$ is a normalizing factor, $m_z$ is the magnetization along easy-axis ($\hat{s}\cdot \vec{m}$ in the present discussion), $\Delta_B=H_kM_s\Omega/2$ is the energy barrier of a magnet with anisotropy field $H_k$, saturation magnetization $M_s$, and volume $\Omega$, $k_B$ is the Boltzmann constant, $T$ is the temperature, $H_{ext}$ is the external magnetic field along the easy-axis, $i_s$ is the $z$-polarized spin current injected into the magnet, and $I_{s0}$ is the critical spin current for magnetization switching \cite{SunPRB2000, SayedEDL2017} for a magnet with PMA, given by
\begin{equation}
\label{SunIs}
I_{s0}=\dfrac{4q}{\hbar}\Delta_B\alpha_g,
\end{equation}
where $\hbar=h/(2\pi)$ and $\alpha_g$ is the Gilbert damping constant.

We consider the case with no external field i.e. $H_{ext}=0$ which from Eqs. \eqref{FPE} and \eqref{SunIs} gives
\begin{equation}
\label{FPE1}
\rho\left(m_z\right)=\dfrac{1}{{\rm Z}}\exp\left(-\dfrac{\Delta_B}{k_B T}\left(1-m_z^2\right)-\left(\dfrac{i_s}{\dfrac{2q}{\hbar}k_B T\alpha_g}\right)m_z\right).
\end{equation}

We consider very low energy barrier magnet i.e. $\frac{\Delta_B}{k_BT}\rightarrow0$, which in Eq. \eqref{FPE1} yields
\begin{equation}
\label{FPE2}
\rho\left(m_z\right)=\dfrac{1}{{\rm Z}}\exp\left(-\dfrac{i_s}{\dfrac{2q}{\hbar}k_B T\alpha_g}m_z\right).
\end{equation}

The steady-state average $\langle m_z \rangle$ is defined as (see Eq. \eqref{av_fpe1})
\begin{equation}
\label{av_fpe}
\langle m_z \rangle = \dfrac{\int_{\phi=-\pi}^{\phi=\pi}\int_{\theta=0}^{\theta=\pi}m_z \;\rho(m_z)\;\sin\theta \;d\theta d\phi}{\int_{\phi=-\pi}^{\phi=\pi}\int_{\theta=0}^{\theta=\pi}\rho(m_z)\;\sin\theta \;d\theta d\phi},
\end{equation}
with $\left(m_z,m_x,m_y\right)\equiv\left(\cos\theta,\sin\theta\cos\phi,\sin\theta\sin\phi\right)$. Combining Eq. \eqref{av_fpe} with Eq. \eqref{FPE2} we get the long time averaged magnetization $\langle m_z \rangle$ for a very low barrier PMA without external magnetic field as
\begin{equation}
\label{langevin}
\langle m_z \rangle = \coth\left(\dfrac{i_s}{\dfrac{2q}{\hbar}k_B T\alpha_g}\right) - \dfrac{\dfrac{2q}{\hbar}k_B T\alpha_g}{i_s},
\end{equation}
which is a Langevin function $L(x)$ of $x\equiv{i_s}/\left({\frac{2q}{\hbar}k_B T\alpha_g}\right)$. 

\begin{figure}
	\includegraphics[width=0.5 \textwidth]{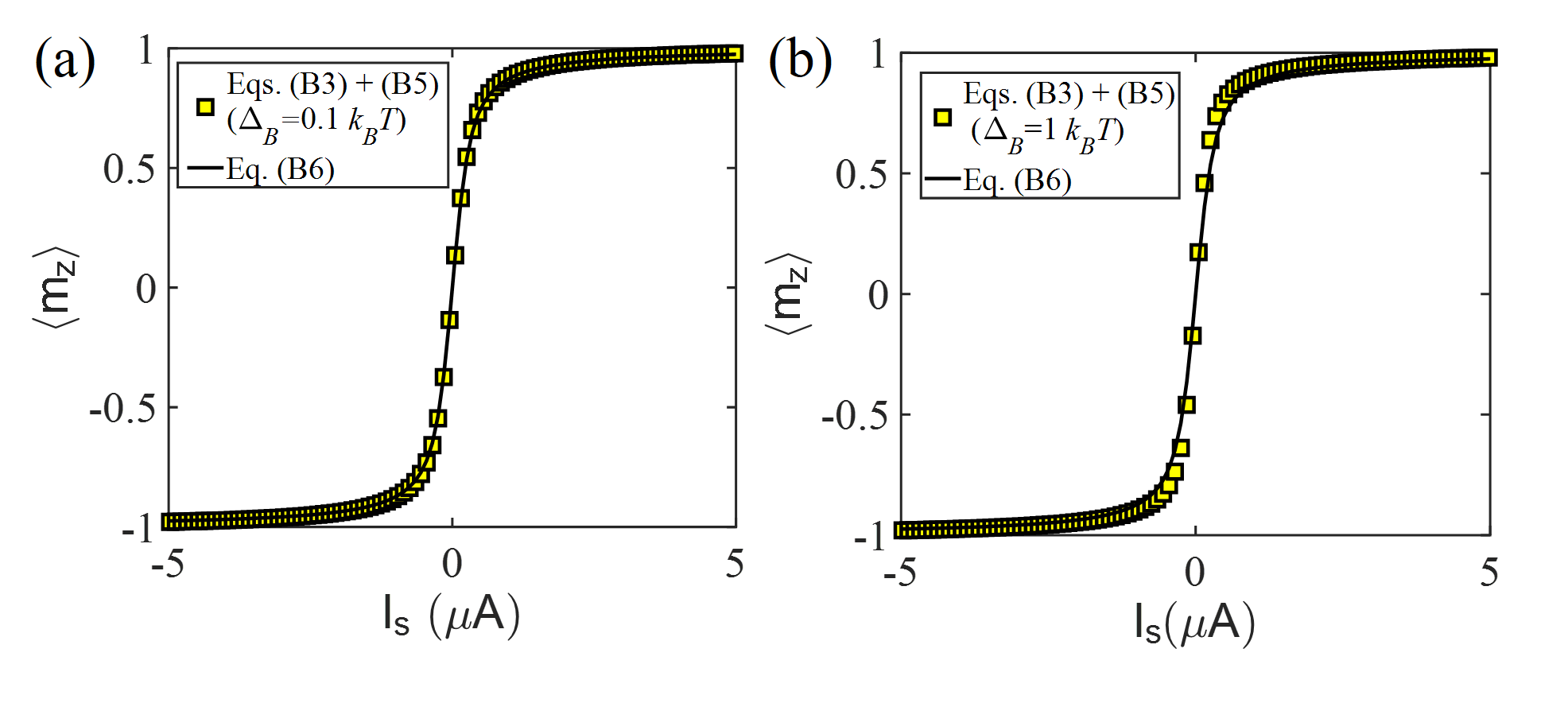}
	\caption{Comparison of simple expression in Eq. \eqref{langevin} which assumes $\Delta_B\rightarrow 0$ with the numerical calculation from Eqs. \eqref{FPE1} and \eqref{av_fpe} for (a) $\Delta_B=0.1k_BT$ and (b) $\Delta_B=1k_BT$. This comparison indicates that  Eq. \eqref{langevin} is reasonably valid for $0\leq\Delta_B\leq1k_BT$.
	}\label{sup1}
\end{figure}

Note that Eq. \eqref{langevin} was derived assuming $\frac{\Delta_B}{k_BT}\rightarrow0$, however, the expression remains reasonably valid up to ${\Delta_B}\approx{k_BT}$. We have compared Eq. \eqref{langevin} with numerical calculations directly from Eqs. \eqref{FPE1} and \eqref{av_fpe} for $\Delta_B=0.1k_BT$ (see Fig. \ref{sup1}(a)) and $k_BT$ (see Fig. \ref{sup1}(b)) respectively, which shows reasonably good agreement. For $\Delta_B>{k_BT}$, the simple expression in Eq. \eqref{langevin} deviates from Eqs. \eqref{FPE1} and \eqref{av_fpe}.

For an estimation of the pinning spin current we can approximate the Langevin function $L(x)\approx \tanh\dfrac{x}{3}$, hence
\begin{equation}
\label{lang_tan}
\langle m_z \rangle \approx \tanh\left(\dfrac{i_s}{\dfrac{6q}{\hbar}k_B T\alpha_g}\right).
\end{equation}

Note that $i_s$ from SML materials are related to input charge current $i_c$ with a conversion factor $\beta$ given by
\begin{equation}
\label{ic_is}
i_s = \beta i_c.
\end{equation}
Combining Eq. \eqref{ic_is} with Eq. \eqref{lang_tan} yields
\begin{equation}
\label{lang_tan_ic}
\langle m_z \rangle \approx \tanh\left(\dfrac{i_c}{\dfrac{6qk_B T\alpha_g}{\hbar \beta}}\right).
\end{equation}

Comparing Eq. \eqref{lang_tan_ic} with Eq. \eqref{avg_Vfm} yields
\begin{equation*}
I_0\approx \dfrac{6qk_B T\alpha_g}{\hbar \beta},
\end{equation*}
which gives the expression in Eq. \eqref{I0_LBM}.

\section{Simulation Setup}
\label{App_Sim}
\noindent\textit{This section provides the details of the simulation setup in SPICE that was used to analyze the proposed rectifier.}\\

\begin{figure}
	\includegraphics[width=0.4 \textwidth]{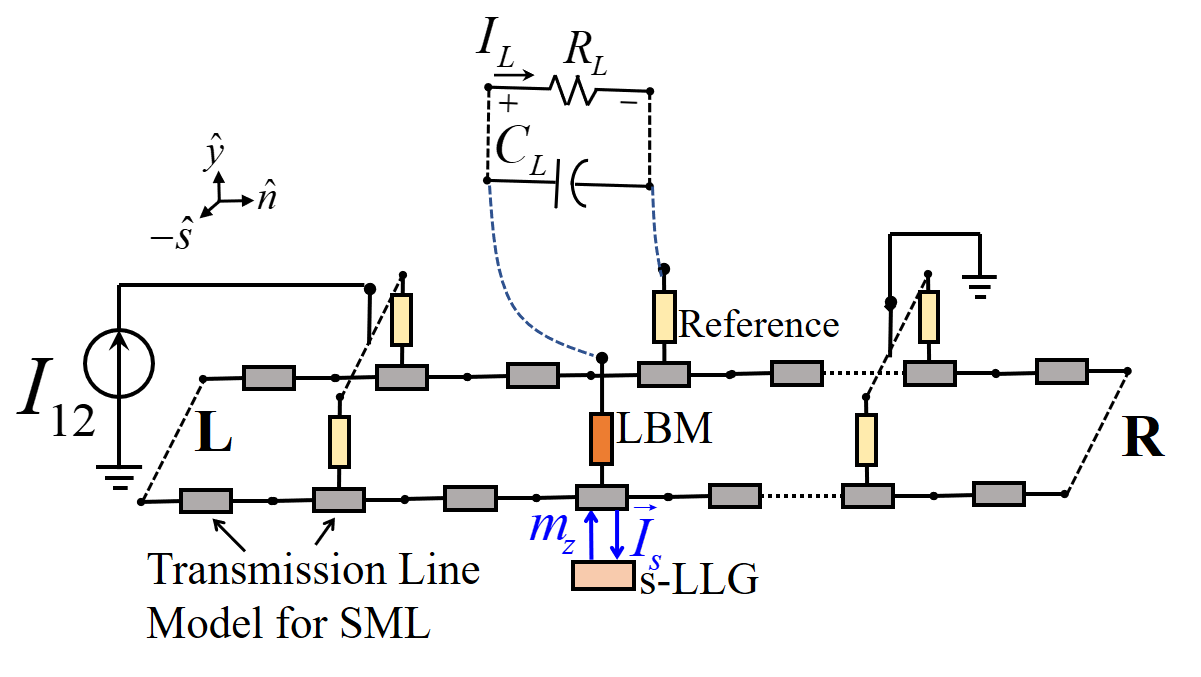}
	\caption{SPICE simulation setup for the structure shown in Fig. \ref{1}(a). SML channel is modeled by connecting SPICE compatible transmission line model \cite{Sayed_arXiv_2017} in a distributed manner. The LBM is modeled with stochastic Landau-Lifshitz-Gilbert (s-LLG) SPICE model \cite{PhysRevX.7.031014}.
	}\label{sup3}
\end{figure}

We have discretized the structure in Fig. {1}(a) into 100 small sections and represented each of the small sections with the corresponding circuit model.  Note that each of the nodes in Fig. \ref{sup3} are two component: charge ($c$) and $z$-component of spin ($s$). We have connected the charge and spin terminals of the models for all the small sections in a modular fashion using standard circuit rules as shown in Fig. \ref{sup3}. The models are connected in a series to reconstruct the structure along length direction. We have two of such parallel chains to take into account the structure along width direction and the two chains represent the area under the LBM and the reference NM respectively. The SML block with LBM is connected to a s-LLG block which takes the spin current from the SML block as input and self-consistently solves for $m_z$ and feeds back to the SML block.

The contacts (1, 2, 3, and 4) in this discussion are point contacts. The polarization of contacts 1, 2, and 4 are $p_f=0$ since they represent normal metals. Polarization of contact 3 is 0.8 which represents an LBM. We set the total number of modes $M+N$ in the channel to be 100. We have assumed that the reflection with spin-flip scattering mechanism is dominant in the channel i.e. $r_{s1,2}\gg r,t_s$. The scattering rate per unit mode was set to 0.04 per lattice point.

We apply the charge open and spin ground boundary condition at the two boundaries given by
\begin{equation}
{\left\{ {\begin{array}{*{20}{c}}
		{{i_c}}\\
		{{v_s}}
		\end{array}} \right\}_L} = \left\{ {\begin{array}{*{20}{c}}
	0\\
	0
	\end{array}} \right\},\,\,\,\,\text{and}\,\,{\left\{ {\begin{array}{*{20}{c}}
		{{i_c}}\\
		{{v_s}}
		\end{array}} \right\}_R} = \left\{ {\begin{array}{*{20}{c}}
	0\\
	0
	\end{array}} \right\}.
\end{equation}
Here, $i_c$ and $v_s$ indicates boundary charge current and boundary spin voltage. Indices $L$ and $R$ indicate left and right boundaries respectively.

Both charge and spin terminals of contact 1 and 2 and the two boundaries of the two parallel model chains are connected together. We apply a current $i_c$ at the charge terminal of contact 1 and make the spin terminal grounded to take into account the spin relaxation process in the contact. We ground both charge and spin terminals of contact 2. The boundary conditions of contacts 1 and 2 are given by
\begin{equation}
{\left\{ {\begin{array}{*{20}{c}}
		{{i_c}}\\
		{{v_s}}
		\end{array}} \right\}_1} = \left\{ {\begin{array}{*{20}{c}}
	i_c\\
	0
	\end{array}} \right\},\,\,\,\,\text{and}\,\,{\left\{ {\begin{array}{*{20}{c}}
		{{v_c}}\\
		{{v_s}}
		\end{array}} \right\}_2} = \left\{ {\begin{array}{*{20}{c}}
	0\\
	0
	\end{array}} \right\}.
\end{equation}
We place a capacitor $C_L$ and load $R_L$ across the charge terminals of contacts 3 and 4. The spin terminals of contacts 3 and 4 are grounded. The boundary conditions of the contacts 3 and 4 are given by
\begin{equation}
{\left\{ {\begin{array}{*{20}{c}}
		{{i_c}}\\
		{{v_s}}
		\end{array}} \right\}_3} = \left\{ {\begin{array}{*{20}{c}}
	0\\
	0
	\end{array}} \right\},\,\,\,\,\text{and}\,\,{\left\{ {\begin{array}{*{20}{c}}
		{{i_c}}\\
		{{v_s}}
		\end{array}} \right\}_4} = \left\{ {\begin{array}{*{20}{c}}
	0\\
	0
	\end{array}} \right\}.
\end{equation}

\section{Sensitivity}
\label{App_ACtoDC}
\noindent\textit{This section discusses the detailed derivation of the sensitivity model in Eq. \eqref{sensitivity}.}\\

We start from Eq. \eqref{spin_vol} with $\vec{m}(t)$ being the instantaneous magnetization of the LBM. and calculate the average as
\begin{equation}
\label{averaging_fm}
\begin{aligned}
\langle V_{34} \rangle &= \dfrac{1}{T} \int_{0}^{T} V_{34}\,dt\\ &=\dfrac{1}{T}\dfrac{{\alpha \xi {p_0}{p_f}}}{{2{G_B}}} \int_{0}^{T} \tanh\left(\dfrac{I_{12}(t)}{I_0}\right)I_{12}(t)\,dt.
\end{aligned}
\end{equation}
Note that the timed average of the random fluctuation in LBM is zero. Here, $G_B=1/R_B$.

We apply an alternating current as input, given by
\begin{equation}
I_{12}(t)=i_{c0}\sin\left(\dfrac{2\pi t}{T} \right),
\end{equation}
The average ac input power applied to the channel with resistance $R_{12}$ is given by
\begin{equation}
\begin{aligned}
P_{in}&=\frac{1}{T}\int_{0}^{T}I_{12}^2(t) R_{12}\,dt\\
&=\frac{i_{c0}^2 R_{12}}{T}\int_{0}^{T}\sin^2 \left(\dfrac{2\pi t}{T}\right)\,dt\\
&=\left(\frac{i_{c0}}{\sqrt{2}}\right)^2R_{12}.
\end{aligned}
\end{equation}

\subsection{Case I: $i_{c0}\gg I_0$}

For $i_{c0}\gg I_0$, we get $\tanh\left(I_{12}(t)/I_0\right)\approx +1$ when $I_{12}(t)>0$ and  $\tanh\left(I_{12}(t)/I_0\right)\approx -1$ when $I_{12}(t)<0$. Thus we have
\begin{equation*}
\tanh\left(\dfrac{I_{12}(t)}{I_0}\right) \times I_{12}(t)\approx |I_{12}(t)|,
\end{equation*}
and from Eq. \eqref{averaging_fm}, we get
\begin{equation}
\label{averaging_fm2}
\begin{aligned}
\langle V_{34} \rangle &=\dfrac{1}{T}\dfrac{{\alpha \xi {p_0}{p_f}}}{{2{G_B}}} \int_{0}^{T} |I_{12}(t)|\,dt\\ &=\dfrac{1}{T}\dfrac{{\alpha \xi {p_0}{p_f}}}{{2{G_B}}} {i}_{c0} \int_{0}^{T} \left|\sin \left(\dfrac{2\pi t}{T}\right)\right|\,dt\\
&=\dfrac{1}{T}\dfrac{{\alpha \xi {p_0}{p_f}}}{{2{G_B}}} {i}_{c0} \int_{0}^{\frac{T}{2}} \sin \left(\dfrac{2\pi t}{T}\right)\,dt\\
&+\dfrac{1}{T}\dfrac{{\alpha \xi {p_0}{p_f}}}{{2{G_B}}} {i}_{c0} \int_{\frac{T}{2}}^{T} -\sin \left(\dfrac{2\pi t}{T}\right)\,dt\\
&=\dfrac{{\alpha \xi {p_0}{p_f}}}{{2{G_B}}} \times  \dfrac{2}{\pi}{i}_{c0}.\\
\end{aligned}
\end{equation}

We write Eq. \eqref{averaging_fm2} as
\begin{equation}
\label{max_vfm2}
\begin{aligned}
\langle V_{34} \rangle=\dfrac{{\alpha \xi {p_0}{p_f}}}{{\pi{G_B}}} \times  \dfrac{\sqrt{2}}{\sqrt{R_{12}}}\times \sqrt{P_{in}},
\end{aligned}
\end{equation}
and the sensitivity is given by
\begin{equation}
\begin{aligned}
\dfrac{d\langle V_{34} \rangle}{dP_{in}} = \dfrac{{\alpha \xi {p_0}{p_f}}}{{\pi{G_B}}} \times  \dfrac{1}{\sqrt{2R_{12}}}\times \dfrac{1}{\sqrt{P_{in}}}.
\end{aligned}
\end{equation}
The sensitivity for $i_{c0}\gg I_0$ decreases inversely proportional to $\sqrt{P_{in}}$. Sensitivity increases for decreasing $P_{in}$ and eventually saturates to a maximum value for $i_{c0}\ll I_0$.

\subsection{Case II: $i_{c0}\ll I_0$}

For $i_{c0}\ll I_0$, we get $\tanh\left(I_{12}(t)/I_0\right)\approx I_{12}(t)/I_0$. Thus from Eq. \eqref{averaging_fm}, we get
\begin{equation}
\label{averaging_fm1}
\begin{aligned}
\langle V_{34} \rangle &=\dfrac{1}{T}\dfrac{{\alpha \xi {p_0}{p_f}}}{{2{G_B}I_0}} \int_{0}^{T} I_{12}^2(t)\,dt\\ &=\dfrac{1}{T}\dfrac{{\alpha \xi {p_0}{p_f}}}{{2{G_B}}} \dfrac{{i}_{c0}^2}{I_0} \int_{0}^{T} \sin^2 \left(\dfrac{2\pi t}{T}\right)\,dt\\
&=\dfrac{{\alpha \xi {p_0}{p_f}}}{{2{G_B}}} \times \dfrac{\left(i_{c0}/\sqrt{2}\right)^2}{I_0} .
\end{aligned}
\end{equation}

We write Eq. \eqref{averaging_fm1} as
\begin{equation}
\label{max_vfm}
\langle V_{34} \rangle =\dfrac{{\alpha \xi {p_0}{p_f}}}{{2{G_B}R_{12}I_0}}  P_{in},
\end{equation}
and the sensitivity is given as
\begin{equation}
\dfrac{d\langle V_{34} \rangle}{dP_{in}}=\dfrac{{\alpha \xi {p_0}{p_f}}}{{2{G_B}R_{12}I_0}},
\end{equation}
which gives the maximum sensitivity in Eq. \eqref{sensitivity}.

\section{Power Conversion Efficiency}
\label{App_Efficiency}
\noindent\textit{This section discusses the ac to dc power conversion efficiency and provides the details of derivation of Eq. \eqref{efficiency}.}\\

Under the no load condition ($R_L\rightarrow\infty$), we have the open circuit dc voltage from Eq. \eqref{max_vfm} for $i_{c0}\ll I_0$
\begin{equation*}
\langle V_{34} \rangle =\dfrac{{\alpha \xi {p_0}{p_f}}}{{2{G_B}R_{12}I_0}}  P_{in},
\end{equation*}
and from Eq. \eqref{max_vfm2} we know that for $i_{c0}\gg I_0$
\begin{equation*}
\begin{aligned}
\langle V_{34} \rangle=\dfrac{{\alpha \xi {p_0}{p_f}}}{{\pi{G_B}}} \times  \dfrac{\sqrt{2}}{\sqrt{R_{12}}}\times \sqrt{P_{in}}.
\end{aligned}
\end{equation*}
Under the short circuit condition ($R_L\rightarrow0$), we have the short circuit dc current $I_L|_{R_L\rightarrow0}=\langle V_{34}\rangle/R_{34}$, where $R_{34}$ is the equivalent resistance between the LBM and the reference NM.

The maximum power transferred to the load is given by
\begin{equation}
P_{L,\max}=\dfrac{1}{4}\times V_L|_{R_L\rightarrow\infty}\times I_L|_{R_L\rightarrow0}=\dfrac{\langle V_{34}\rangle^2}{4R_{34}}.
\end{equation}
which yields
\begin{equation}
\begin{aligned}
P_{L,\max}&=\left(\dfrac{{\alpha \xi {p_0}{p_f}}}{{2{G_B}R_{12}I_0}} \right)^2 \dfrac{P_{in}^2}{4R_{34}}\;\;\;\text{for}\;\;i_{c0}\ll I_0\\
&=\left(\dfrac{{\alpha \xi {p_0}{p_f}}}{{\pi{G_B}}}\right)^2  \dfrac{P_{in}}{{2R_{12}R_{34}}} \;\;\;\text{for}\;\;i_{c0}\gg I_0.
\end{aligned}
\end{equation}

The ac to dc power conversion efficiency is given by
\begin{equation}
\begin{aligned}
\eta=\dfrac{dP_{L,\max}}{dP_{in}}&=\left(\dfrac{{\alpha \xi {p_0}{p_f}}}{{2{G_B}R_{12}I_0}} \right)^2 \dfrac{P_{in}}{2R_{34}}\;\;\;\text{for}\;\;i_{c0}\ll I_0\\
&=\left(\dfrac{{\alpha \xi {p_0}{p_f}}}{{\pi{G_B}}}\right)^2  \dfrac{1}{{2R_{12}R_{34}}} \;\;\;\text{for}\;\;i_{c0}\gg I_0.
\end{aligned}
\end{equation}
Note that $\eta$ increases with input ac power $P_{in}$ and reaches a maximum when $i_{c0}\gg I_0$ given by
\begin{equation*}
\begin{aligned}
\eta_{\max}=\left(\dfrac{{\alpha \xi {p_0}{p_f}}}{{\pi{G_B}}}\right)^2  \dfrac{1}{{2R_{12}R_{34}}}.
\end{aligned}
\end{equation*}

\bibliography{TI_rect,Ref,Ref1,Ref2}

\end{document}